\documentclass[a4paper,fleqn,usenatbib]{mn2e}

\usepackage[T1]{fontenc}
\usepackage{ae,aecompl}

\usepackage{graphicx}    
\usepackage{amsmath}    
\usepackage{amssymb}    

\usepackage{ulem}

\newcommand{\etal}{{et al.}}
\def\apj{{ApJ}}%
\def\aap{{A\&A}}%
\def\mnras{{MNRAS}}%
\def\na{{New Astron.}}%
\def\pasj{{PASJ}}%
\title[Relativistic inviscid accretion flow]{Low angular momentum relativistic hot accretion flow around Kerr black holes with variable adiabatic index}

\author[Dihingia et al.]{
Indu K. Dihingia$^{1}$\thanks{E-mail: i.dihingia@iitg.ernet.in},
Santabrata Das$^{1}$\thanks{E-mail: sbdas@iitg.ernet.in},
Anuj Nandi$^{2}$\thanks{E-mail: anuj@isac.gov.in}
\\
$^{1}$Indian Institute of technology Guwahati, Guwahati, 781039, India\\
$^{2}$Space Astronomy Group, ISITE Campus, U. R. Rao Satellite Center, Outer Ring Road, Marathahalli, Bangalore, 560037, India
}

\date{Accepted XXX. Received YYY; in original form ZZZ}

\pubyear{2016}

\begin{document}
\label{firstpage}
\pagerange{\pageref{firstpage}--\pageref{lastpage}}
\maketitle

\begin{abstract}
	We study the relativistic, time-independent, low angular momentum, inviscid, advective accretion flow around Kerr black hole. Considering the relativistic equation of state (REoS), we examine the transonic properties of the flow and find that there exists an upper bound of the location of the physically accepted critical point ($r^{\rm max}_{\rm out}$). However, no such limit exists when an ideal gas equation of state (IEoS) is assumed to describe the flow. Further, we calculate the global accretion solutions that contain shock waves and separate the domain of parameter space in angular momentum ($\lambda$) and energy (${\cal E}$) plane. We find ample disagreement between the shock parameter spaces obtained for REoS and IEoS, respectively. In general, post-shock flow (equivalently post-shock corona, hereafter PSC) is characterized by shock location ($r_s$) and compression ratio ($R$, measure of density compression across the shock front) which are uniquely determined for flow with given input parameters, namely $({\cal E}, \lambda)$. Using $r_s$ and $R$, we empirically compute the oscillation frequency ($\nu_{QPO}$) of the shock front which is in general quasi-periodic (QP) in nature and retrace the domain of shock parameter space in the $r_s-R$ plane in terms of $\nu_{QPO}$ for REoS around the weakly as well as rapidly rotating black holes. Finally, we indicate the relevance of the present work to explain the plausible origin of high frequency QPO (HFQPO) and its connection with the spin ($a_k$) of the Galactic black hole sources.
	
\end{abstract}

\begin{keywords}
	accretion, accretion discs - black hole physics - hydrodynamics - shock waves.
\end{keywords}

\clearpage

\section{Introduction}

Accretion of gas onto a black hole is widely considered to be the underlying source of energy in most energetic astrophysical objects in the universe, namely X-ray binaries (XRBs) and active galactic nuclei (AGNs) \citep{Frank-etal02}. In general, these objects often exhibit spectral and temporal variabilities in X-rays that eventually carry the imprints of the physical processes active in the accretion flow surrounding the black holes. In particular, the spectral state transitions from low/hard state to high/soft state via intermediate states are observed in black hole XRBs  and these sources also display the signature of Quasi-periodic Oscillations (QPOs) as well \citep{Belloni-etal2002,Chakrabarti-etal2002,Homan-Belloni2005,Remillard-McClintock2006,Nandi-etal2012,Iyer-etal2015,Nandi-etal2018}.


So far, several efforts were made to understand the origin of the above mentioned X-ray variabilities. \citet{Tagger-Pellat1999} examined the accretion-ejection instability and pointed out that such instability can explain the QPO features. \citet{Titarchuk-Osherovich2000} suggested the global disc oscillation mechanism that possibly causes the persistent QPOs observed in the black hole systems. \citet{Ingram-Done2011} studied the evolution of X-ray timing properties ($i.e.$, QPOs) by means of the fluctuations associated with the hot flows at the inner part of the disc.  On the other hand, \cite{Svensson-Zdziarski1994,Esin-etal1997,Done-Kubota2006} studied the spectral properties of the accretion flow considering Compton corona coupled with the standard Keplerian Disc \citep{ss73}. 

Adopting a self-consistent approach, Chakrabarti and his collaborators \citep[and references therein]{Chakrabarti-Titarchuk1995,Mandal-Chakrabarti2005a,Mandal-Chakrabarti2005b,Nandi-etal2012,Debnath-etal2014,Iyer-etal2015,Nandi-etal2018} have also been investigating both spectral and temporal properties of the Galactic black hole sources since more than two decades where they particularly examine the importance of shock wave in accretion flow.
This accretion flow model is developed based on the solution of Two Component Advective Flow (TCAF) paradigm that invokes the role of the post-shock flow ($i.e.$, PSC, equivalent to `Compton corona'), where due to shock compression, density and temperature become relatively higher compared to the pre-shock flow. In reality, soft photons from the pre-shock flow are intercepted by the hot electrons at PSC and are reprocessed via inverse Comptonization mechanism to produce hard radiations. When PSC modulates, the emergent hard radiations also oscillate that eventually exhibits QPO features \citep{Das-etal14,Sukova-Janiuk15,Lee-etal2016}. Interestingly, the same PSC region deflects a part of the inflowing matter in the transverse direction of the flow motion to originate bipolar outflows \citep{Chakrabarti1999,Das-etal01,Das-Chattopadhyay08,Aktar-etal2015,Aktar-etal2017}. In the absence of PSC, accretion flow behaves similarly as in the standard thin disc \citep{ss73}, where outflow generation ceases to exist. Overall, in this model, PSC plays a vital role in determining the emergent flux of the high energy radiations as well as the mass outflow rate from the disc \citep{Chakrabarti-Titarchuk1995,Nandi-etal2018}. 


Meanwhile, the study of the shock waves in the accretion flow around 
black holes is being carried out by the numerous group of workers both 
theoretically as well as numerically \citep{Fukue1987,Chakrabarti1989,Yang-Kafatos1995,Molteni-etal96,
	Ryu-etal1997,Lu-etal99,Becker-Kazanas01,Fukumura-Tsuruta2004,
	Das2007,Kumar_etal2013,Das-etal14,Okuda-Das2015,Sukova-Janiuk15,
	Sarkar-Das16,Aktar-etal2017,Dihingia-etal2018}. 
Very recently, \cite{Kim-2017,Kim-2018} showed the formation of shocks in the accretion flow around black holes using general relativistic hydrodynamical numerical simulation. 
In addition, \cite{Nishikawa-2005} and \cite{Fukumura-2016} also examined the shock solutions in the GRMHD framework under suitable physical conditions.
In general, in an accretion disc, rotating matter experiences 
centrifugal barrier while falling towards the black hole. Depending 
on the flow parameters, centrifugal repulsion becomes strong enough 
to trigger the discontinuous transition of the flow variables in the 
form of shock waves. Since PSC is formed because of the shock transition, 
it is generally characterized by the shock variables, namely shock 
location ($r_s$) and compression ratio ($R$). In a way, both $r_s$ and $R$ are interrelated and 
also associated with a given accretion solution that harbors shock wave 
\citep{Das-etal01b,Das2007,Sarkar-Das16,Chattopadhyay-Kumar2016,Aktar-etal2017,
	Dihingia-etal2018}. 
In addition, in an 
accretion disc, the flow is expected to be in the thermally relativistic domain 
($i.e.$, adiabatic index $\Gamma \rightarrow 4/3$) at the inner part of the disc 
whereas it remains thermally non-relativistic ($i.e.$, $\Gamma \rightarrow 5/3$) 
at a distance far away from the black hole horizon \citep{Frank-etal02}. Nonetheless, for
simplicity, $r_s$ and $R$ are generally computed for accretion flows obeying
an ideal equation of state (hereafter IEoS), where the value of the adiabatic 
index ($\Gamma$) remains constant all throughout the flow
\citep[and references 
there in]{Abramowicz-Chakrabarti1990,Yang-Kafatos1995,Chakrabarti1996,Lu-etal99,
	Das-etal01b,Fukumura-Tsuruta2004,Chakrabarti-Das2004,Das2007,Sarkar-Das16,
	Aktar-etal2017,Dihingia-etal2018}. Recently, \citet{Kumar_Chattopadhyay2017} 
studied the accretion-ejection solutions using general relativistic prescription;
however, no attempts were made to compute the limiting range of $r_s$ and $R$ 
and their correlation for the relativistic accretion flows around rotating 
black holes.

Being motivated with these, in the present work, we aim to study the structure of the 
relativistic accretion flows around the Kerr black holes, where inflowing matter is described with a relativistic equation of state (REoS). Here, adiabatic index ($\Gamma$) of the flow no longer remains fixed as it was the case in many earlier works; instead $\Gamma$ is determined here self-consistently based on the thermal properties of the flow \citep{Chattopadhyay_Ryu2009}. To begin with, we consider low angular momentum inviscid transonic accretion flow in the steady state and examine its transonic properties considering both REoS and IEoS. Further, we calculate the shock induced global accretion solutions around rotating black holes and identify the effective region of the parameter space in energy (${\cal E}$) and angular momentum ($\lambda$) plane that permits shock solutions. On comparing the shock parameter spaces obtained for REoS and IEoS, ample disagreement is seen. Further, we empirically  compute the frequency of QPO ($\nu_{\rm QPO}$) of the shock front \citep{Chakrabarti-Manickam00,Iyer-etal2015}. Since $r_s$ and $R$ are uniquely determined for a shocked accretion flow having fixed ($\lambda, {\cal E}$), we identify the shock parameter space in $r_s$-$R$ plane in lieu of the canonical $\lambda-{\cal E}$ plane where we study the two-dimensional projection of three-dimensional plot of $\{r_s, R, \nu_{QPO}\}$. With this, in this work, we study the importance of REoS over IEoS while obtaining the accretion solutions and also investigate the $r_s - R$ correlation for shocked accretion flow around Kerr black holes. Finally, we discuss the implication of the present formalism to study the HFQPO and its possible association with the spin of the Galactic black hole sources.

The paper is organized as follows. In \S 2, we present the governing equations of the relativistic accretion flow. In \S 3, we discuss the critical point analysis. In \S 4, we present the results, where critical point properties, global accretion solutions including shocks, shock properties are discussed. In \S 5, we discuss the astrophysical implication of our formalism. Finally, in \S 6, we present the concluding remarks.    

\section{Model Equations and Assumptions}

In the present work, the accretion disc around a rotating black hole is considered to be steady, thin, axisymmetric and non-dissipative in nature. Assumption of inviscid accretion flows around the black hole is supplemented in appendix A.The black hole is characterized by its mass $M_{\rm BH}$ and spin $a_k=J/ M_{\rm BH}$, where $J$ is the angular momentum of the black hole. Throughout the study, we use a unit system as $G=M_{\rm BH}=c=1$, where $G$ and $c$ are the gravitational constant and speed of light. In this unit system, length, angular momentum and time are expressed in terms of $GM_{\rm BH}/c^2$, $GM_{\rm BH}/c$ and $GM_{\rm BH}/c^3$, respectively.

\subsection{Equations of the fluid}
The non-dissipative energy momentum tensor for fully ionized fluid is expressed in terms of energy density $(e)$, pressure $(p)$ and four velocities $(u^\mu)$ and is given by,
$$
T^{\mu\nu} = (e+p)u^\mu u^\nu + pg^{\mu\nu},
\eqno(1)
$$
where $\mu$ and $\nu$ are indices run from $0 \to 3$, $g^{\mu\nu}$ are the components of the metric. The conservation of mass flux and the conservation of energy momentum tensor constitute the governing equations of hydrodynamics which are given by,
$$
T^{\mu\nu}_{;\nu}=0,\qquad (\rho u^\nu)_{;\nu}=0,
\eqno(2)
$$ 
where $\rho$ is the mass density of the flow. Now, we define the projection 
operator $h^i_\mu = \delta^i_\mu + u^i u_\mu$ that satisfy $h^i_{\mu}u^{\mu}=0$ with  `$i$' runs from $1 \to 3$. This condition helps us to project the Navier-Stokes equation into three vector equations as,
$$
h^i_\mu {T_0^{\mu\nu}}_{;\nu}=(e+p)u^\nu u^i_{;\nu} + (g^{i\nu} + u^i u^\nu)p_{,\nu}=0. 
\eqno(3)
$$
In addition, the scalar equation which is essentially identified as the first law of thermodynamics is computed as,
$$
u_\mu T^{\mu\nu}_{;\nu}=u^\mu\bigg[\left(\frac{e+p}{\rho}\right)\rho_{,\mu} - e_{,\mu}\bigg]=0.
\eqno(4)
$$

In this work, we intend to study the accretion flow around a Kerr black hole and therefore, we chose Kerr metric in Boyer-Lindquist coordinates as,
$$\begin{aligned}
ds^2 = &g_{\mu\nu} dx^\mu dx^\nu,\\
     = & g_{tt}dt^2 + 2g_{t\phi}dtd\phi + g_{rr} dr^2 + g_{\theta\theta} 
     d\theta^2 + g_{\phi\phi} d\phi^2,
\end{aligned}\eqno(5)$$
where $x^\mu  (\equiv t,r,\theta,\phi)$ denote coordinates and
$g_{tt} = -(1 - 2r/\Sigma)$, $g_{t\phi} = -2a_kr\sin^2\theta/\Sigma$, 
$g_{rr}=\Sigma/\Delta$, $g_{\theta\theta}=\Sigma$ and $g_{\phi\phi}=A\sin^2\theta/\Sigma$ 
are the non-zero metric components. Here, $A=(r^2 + a_k^2)^2 - \Delta a_k^2
\sin^2\theta$, $\Sigma = r^2 + a_k^2\cos^2\theta$ and $\Delta = r^2 - 2r + a_k^2$.
In this work, we follow a convention where the four velocities satisfy $u_\mu u^\mu=-1$.

To obtain the accretion solutions, one requires to use the equation of state (EoS) describing the relation among the thermodynamical quantities, namely density ($\rho$), pressure ($p$) and internal energy ($e$), respectively. Since the temperature of the accretion flow generally exceeds $\sim 10^{10}$K at least within few tens of Schwarzschild radius \citep[and references therein]{Sarkar-Das16}, in this work, we consider a simplified EoS for relativistic fluid (hereafter REoS) consisting of electrons, positrons and ions \citep{Chattopadhyay_Ryu2009} and is given by,
$$
e = n_em_ef=\frac{\rho}{\tau}f.
\eqno(6)
$$
Here, $\rho = n_em_e\tau$, $\tau = [2 - \zeta(1 - 1/\chi)]$, $\zeta = n_p/n_e$ 
and $\chi=m_e/m_p$, respectively, where $n_i$'s and $m_i$'s are the number density and the mass of the species. Throughout this study, we consider the flow to be composed with electrons and ions only and we set $\zeta=1$, until otherwise stated. Subsequently, the explicit form of $f$ is obtained as,
$$
f = (2-\zeta)\bigg[1 + \Theta\left(\frac{9\Theta + 3}{3\Theta + 2}\right)\bigg] +
 \zeta\bigg[ \frac{1}{\chi} + \Theta\left(\frac{9\Theta + 3/\chi}
{3\Theta + 2/\chi}\right)\bigg],
\eqno(7)
$$
where $\Theta~(= k_{\rm B}T/m_ec^2$) is the dimensionless temperature. In this context, we define the polytropic index $(N)$, adiabatic index $(\Gamma)$ and the sound 
speed $(a_s)$ as,
$$
N = \frac{1}{2}\frac{df}{d\Theta}; \quad \Gamma = 1 + \frac{1}{N}; {\rm and}
\quad a_s^2 = \frac{\Gamma p}{e+p} = \frac{2\Gamma\Theta}{f + 2\Theta}.
\eqno(8)
$$
The essence of REoS is that during accretion, the flow variables determines the $\Gamma$ variation as expected. 

It may be noted that because of simplicity, the EoS widely used in the literature is described with a fixed adiabatic index $\Gamma$ (ideal EoS, hereafter IEoS) and is given by,
$$
e=\frac{p}{\Gamma -1} + \rho=\frac{\rho}{\tau} f,
\eqno(9)
$$
where $f=2N\Theta + \tau$. For the purpose of completeness, it would be worthy to compare results obtained for both REoS and IEoS, respectively.

\subsection{Governing Equations for Accretion Disc}

In this work, since a geometrically thin accretion disc is assumed, it is justified to consider the accreting matter to be confined at the disc equatorial plane. Accordingly, we choose $\theta=\pi/2$ and $u^\theta\sim 0$. Using these conditions, the radial component of the equation (3) takes the form as,
$$
\begin{aligned}
&u^ru^r_{,r} + \frac{1}{2}g^{rr}\frac{g_{tt,r}}{g_{tt}} + \frac{1}{2}u^ru^r
\left(\frac{g_{tt,r}}{g_{tt}} + g^{rr}g_{rr,r}\right)\\
 + &u^\phi u^tg^{rr}\left(\frac{g_{t\phi}}{g_{tt}}
g_{tt,r} - g_{t\phi,r}\right)
+\frac{1}{2}u^{\phi}u^{\phi}g^{rr}\left(\frac{g_{\phi\phi}g_{tt,r}}{g_{tt}} - g_{\phi\phi,r}\right)\\
&+ \frac{(g^{rr} + u^ru^r)}{e+p}p_{,r}=0.\\
\end{aligned}
\eqno(10)
$$

In addition, the continuity equation (2nd part of equation 2) can be rewritten as the mass accretion rate which is given by,
$$
\dot{M} = -4\pi r u^r \rho H,
\eqno(11)
$$
where $\dot{M}$ represents the accretion rate that we treat as global constant. Moreover, $H$ refers the local half-thickness of the disc which has the functional
form \citep{Riffert_Herold1995,Peitz_Appl1997} as,
$$
H^2 = \frac{pr^3}{\rho \mathcal{F}},
\eqno(12)
$$
with
$$
\mathcal{F}=\gamma_\phi^2\frac{(r^2 + a_k^2)^2 + 2\Delta a_k^2}
{(r^2 + a_k^2)^2 - 2\Delta a_k^2},
$$
where $\gamma_\phi^2=1/(1-v_\phi^2)$ and $v_\phi^2 = u^\phi u_\phi/(-u^t u_t)$. We define the radial three velocity in the co-rotating frame as $v^2 = \gamma_\phi^2v_r^2$ and thus we have $\gamma^2 = 1/(1-v^2)$, where $v_r^2 = u^ru_r/(-u^tu_t)$. 

We adopt a stationary metric $g^{\mu\nu}$ which has axial symmetry. This enables us to 
construct two Killing vector fields $\partial_t$ and $\partial_\phi$ that provide two conserved quantities for the fluid motion in this gravitational filed and are given by, 
$$
hu_\phi = {\rm constant}; \qquad -hu_t={\rm constant} = {\cal E},
\eqno(13)
$$
where $h~[=(e+p)/\rho]$ is the specific enthalpy of the fluid, ${\cal E}$ is the Bernoulli constant ($i. e.$, the specific energy of the flow) and 
$u_t = -\gamma \gamma_\phi/\sqrt{g^{t\phi}\lambda - g^{tt}}$, where
$\lambda ~(= -u_\phi/u_t$) denotes the conserved specific angular momentum.

\section{Critical Point Analysis}

Simplifying equations (4), (6), (10) and (11), we obtain the wind equation in the co-rotating frame as,
$$
\frac{dv}{dr}= \frac{\mathcal{N}}{\mathcal{D}},
\eqno(14)
$$
where the numerator ${\cal N}$ is given by,
$$\begin{aligned}
\mathcal{N}=& - \frac{1}{(r-2) r} + \gamma_\phi^2\frac{2a_k}{r^2\Delta}\lambda 
+ \gamma_\phi^2\frac{4a_k^2}{(r-2)r^2\Delta}\\
& -\gamma_\phi^2\Omega\lambda\frac{ 2a_k^2- (r-3) r^2}{r^2\Delta}
 + 2a_k\gamma_\phi^2\Omega\frac{ (r-3) r^2-2 a_k^2}{(r-2) r^2 \Delta}\\
&+ \frac{2a_s^2}{\Gamma + 1}\bigg[ \frac{\left(r-a_k^2\right)}{r\Delta} 
+ \frac{5}{2r} - \frac{1}{2\mathcal{F}}\frac{d\mathcal{F}}{dr}\bigg],\\
\end{aligned}\eqno(15)$$
and the denominator ${\cal D}$ is given by,
$$
\mathcal{D} = \gamma^2\bigg[v- \frac{2a_s^2}{v(\Gamma +1)}\bigg].
\eqno(16)
$$

Using equation (14) in equation (4), we calculate the derivative of the dimensionless temperature as,
$$
\frac{d\Theta}{dr}=-\frac{2\Theta}{2N + 1}\bigg[\frac{\left(r-a_k^2\right)}{r\Delta}
+\frac{\gamma^2}{v}\frac{dv}{dr} + \frac{5}{2r} - 
\frac{1}{2\mathcal{F}}\frac{d\mathcal{F}}{dr}\bigg].
\eqno(17)$$ 

In case of accretion process around the black holes, since the inflowing matter smoothly accretes all throughout starting from the outer edge of the disc ($i.e.,$ subsonic) up to the horizon ($i.e.,$ suppersonic), the radial velocity gradient (equation 14) must be real and finite always. However, equation (16) indicates that the denominator (${\cal D}$) may vanish at some points and therefore, to keep $dv/dr$ finite, numerator (${\cal N}$) must also goes to zero there. These spatial points where both ${\cal N}$ and ${\cal D}$ simultaneously tend to zero ($i. e.,$ $dv/dr=0/0$) are called as critical points and the conditions ${\cal N}=0$ and ${\cal D}=0$ are known as critical point conditions. Using condition ${\cal D}=0$, we obtain the expression of Mach number ($M=v/a_s$) at the critical point ($r_c$) as,
$$
M_c \equiv M(r_c) = \sqrt{\frac{2}{\Gamma_c + 1}}, 
\eqno(18)
$$

Setting ${\cal N}=0$, we have
$$
v_c^2 = \frac{\mathcal{N}_N}{\mathcal{N}_D},
\eqno(19)
$$
with
$$\begin{aligned}
\mathcal{N}_N =& \left[ \frac{1}{(r-2) r} - \frac{2a_k\gamma^2_{\phi}\lambda}{r^2\Delta}
+\frac{\gamma_{\phi}^2\Omega\lambda \left[ 2a_k^2- (r-3) r^2 \right]}{r^2\Delta}\right]_c\\
& - \left[\frac{4a_k^2\gamma^2_{\phi}}{(r-2)r^2\Delta} 
+ \frac{ 2a\gamma_{\phi}^2\Omega \left[(r-3) r^2-2 a_k^2\right]}{(r-2) r^2 \Delta}\right]_c,
\end{aligned}$$
and
$$
\mathcal{N}_D = \left[\frac{\left(r-a_k^2\right)}{r\Delta} + \frac{5}{2r} 
- \frac{1}{2\mathcal{F}}\frac{d\mathcal{F}}{dr}\right]_c,
$$
where subscript `$c$' denotes the quantities evaluated at the critical point ($r_c$). By solving equations (18) and (19) with help of equation (8), we calculate $\Theta_c$ and $v_c$ at the critical point. These values serve as the initial conditions to 
integrate equations (14) and (17). Before integrating these equations, the values of 
$dv/dr|_c$ need to be determined that eventually provide the essence of the critical point characteristics. It is noteworthy that $dv/dr|_c$ usually assumes two values: when both values of $dv/dr|_c$ are real and of opposite sign, critical points are known as saddle type; if $dv/dr|_c$ takes two value of same sign, critical points are known as nodal type; and the spiral-type critical point is obtained when $dv/dr|_c$ becomes imaginary \citep{Holzer1977}.  To calculate $dv/dr|_c$, we employ l\'Hospital rule and the 
explicit expression is obtained as,
$$
\frac{dv}{dr}\bigg|_c=\frac{N_1-D_2\pm\sqrt{(N_1-D_2)^2+4D_1N_2}}{2D_1},
\eqno(20)
$$
where $N_1,N_2,D_1$ and $D_2$ are the functions of the flow variables which are given in the appendix B. It may be noted that saddle type critical points are specially important in accretion disc as they are stable \citep{kato_etal1993} and accretion flow smoothly passes through it \citep{Liang_Thompson1980,Abramowicz_Zurek1981,Chakrabarti1989} before entering into the black hole. Thus, in a realistic scenario, accretion flow must contain at least one saddle type critical point. In the subsequent sections, we refer all the saddle type critical points as critical points unless otherwise stated.

\section{Results and Discussions}

\subsection{Nature of Critical Points}

\begin{figure}
	\includegraphics[scale=0.45]{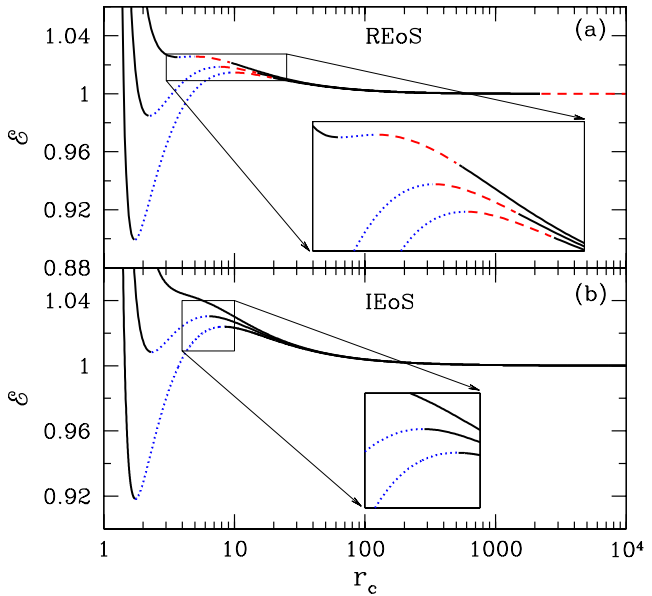}
	\caption{Plot of energy at the critical point $({\cal E})$ as a function of critical 
		point location $(r_{\rm c})$ for three angular momentum $\lambda = 2.00, 1.90$ and $1.80$ (bottom to top) in both the panels with $a_k=0.99$.
		Here, results presented in the upper and lower panels are for the 
		REoS and IEoS, respectively. For IEoS, we choose $\Gamma=1.4$. Solid, dash and dotted curves represent results corresponding to the saddle, nodal and spiral-type critical points. See text for details.}
	\label{fig:fig01}
\end{figure}

In order to calculate the location of critical points, 
we solve the second part of equation (13) by supplying the global parameters ${\cal E}$, $\lambda$ and $ a_k$, respectively. Depending on the choice of the parameters, the flow may contain either single or multiple critical points through which it enters into the black hole \citep{Fukue1987,Chakrabarti1989}. In this section, we investigate the transonic nature of the accretion flow and in Fig. \ref{fig:fig01}, we present the variation of the flow energy (${\cal E}$) as function of logarithmic critical point locations ($r_c$) for various angular momentum around a black hole having spin $a_k=0.99$.
The obtained results are plotted for REoS and IEoS in the upper panel (Fig. 1a) and lower panel (Fig. 1b), respectively and in each panel, various curves from top to bottom are for different angular momentums which are given by $\lambda = 1.8, 1.9$ and $2.0$, respectively. In case of IEoS, we consider $\Gamma = 1.4$ as a representative value. The solid, dashed and dotted parts of the curve represent the saddle type, nodal type, and spiral-type critical points. We observe that for REoS, all three types of critical points are present in systematic order: for example, saddle --- spiral --- nodal --- saddle --- nodal as the critical point locations are increased. On the contrary, nodal type critical point is absent for IEoS. Moreover, we observe that for a given angular momentum, there exists a range of energy  
that provides multiple critical points in the flow. Among them, the closest one from the black hole is called as the inner critical point ($r_{\rm in}$), and the furthest one is called as the outer critical point ($r_{\rm out}$). Since only saddle type critical points are physically acceptable, we find an upper limit of outer critical point ($r^{\rm max}_{\rm out}$) for REoS whereas $r_{\rm out}$ remains unbounded for IEoS. Usually, since accretion flow encounters discontinuous shock transition in between $r_{\rm in}$ and $r_{\rm out}$ \citep{Chakrabarti1989}, the maximum possible shock radius ($r^{\rm max}_s$) will have an upper bound for REoS. In case of IEoS, since $r^{\rm max}_{\rm out}$ is practically limitless, $r^{\rm max}_s$ also becomes unbounded. In reality, the accretion flow remains thermally non-relativistic ($i.e.$, $\Gamma \rightarrow 5/3$) at far away distances from the black hole as the temperature of the accretion flow is small. When the flow proceeds towards the black hole, its temperature increases due to compression and eventually flow becomes thermally relativistic ($i.e.$, $\Gamma \rightarrow 4/3$) at the inner part of the disc. Note that REoS describes the above features of the accretion flow very much satisfactorily and thus we do not find any transonic accretion solution (absence of saddle type critical points) in the non-relativistic regime. These findings are consistent with results reported in \cite{Chakrabarti1990}. In case of IEoS, as $\Gamma$ is considered to be constant (and $4/3 \le \Gamma < 5/3$), $r_{\rm out}$ continues to exist even at the outer edge of the disc. In the inset, we zoom a small part of the curves for clarity purposes only. 

\subsection{Parameter Space for Multiple Critical Points}

\begin{figure}
	\includegraphics[scale=0.425]{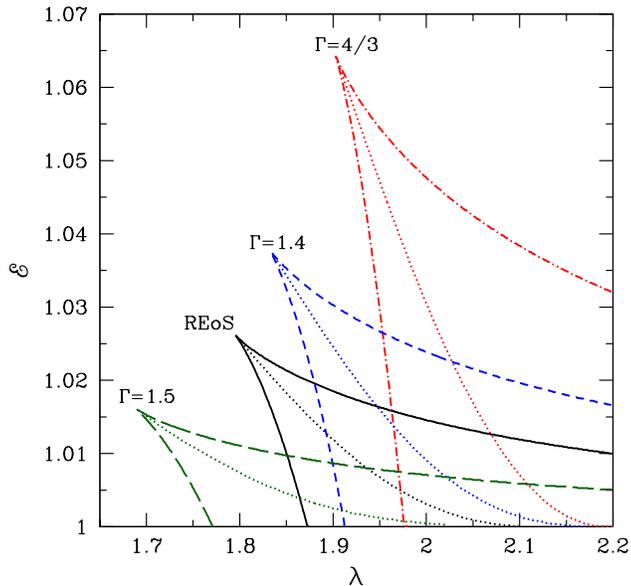}
	\caption{Comparison of parameter spaces for multiple critical points in $\lambda-{\cal E}$ plane. Region bounded by the solid curve is for REoS and the region bounded by dot-dashed, small-dash and long-dash curves are for IEoS with $\Gamma = 4/3$, $1.4$ and $1.5$, respectively. The domain of each parameter space is further subdivided using dotted curve that corresponds to accretion solutions having equal entropy at the inner and outer critical points. See text for details.
	}
	\label{fig:fig02}
\end{figure}

In this section, we study the parameter space in $\lambda-{\cal E}$ plane for the accretion flow that contains multiple critical points. In Fig. \ref{fig:fig02}, we identify the effective region of the parameter space bounded by the solid curves for REoS whereas the region separated using dot-dashed ($\Gamma = 4/3$), dashed ($\Gamma = 1.4$) and long-dashed ($\Gamma = 1.5$) are obtained for IEoS. Note that all four  parameter spaces are further subdivided based on the ratio of entropies ($\eta = {\dot {\cal M}}_{\rm in}/{\dot {\cal M}}_{\rm out}$) measured at inner and outer critical points. Dotted curve in every parameter space corresponds to $\eta = 1$: above and below the dotted curve we have $\eta < 1$ and $\eta > 1$, respectively. It is also clear that the multiple critical point parameter space for REoS only display a partial overlap with the remaining cases and hence, we point out that any observable computed using IEoS is expected to be erroneous. 

\subsection{Global Accretion Solution with Shock}

\begin{figure}
	\includegraphics[scale=0.425]{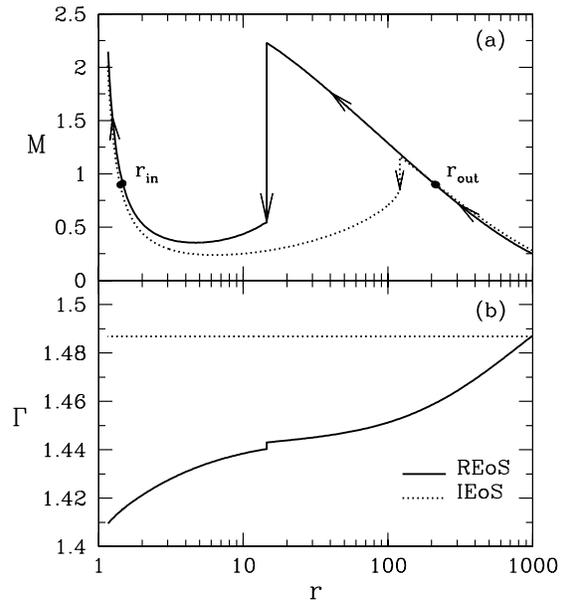}
	\caption{Comparison of accretion solutions containing shock waves for REoS and IEoS. In the upper panel (a), variation of Mach number is presented as function of radial coordinate ($r$). Here, input parameters are chosen as ${\cal E}= 1.001, \lambda=1.98$ and $a_k=0.99$, respectively. In both cases, flows are injected at $r=10^{3}$ with adiabatic index $\Gamma = 1.4896$. Solid and dashed curves are used to depict the results corresponding to REoS and IEoS. In the lower panel (b), variation of $\Gamma$ is plotted with $r$. See text for details.
	}
	\label{fig:fig03}
\end{figure}

In the previous section, we have shown that depending on the flow parameters, such as energy (${\cal E}$) and angular momentum ($\lambda$), accretion flow may possess both inner ($r_{\rm in}$) and outer ($r_{\rm out}$) critical points. 
Interestingly, flow can not smoothly pass through both critical points simultaneously unless it makes a transition in between them. In actuality, inflowing matter from the outer edge of the disc first crosses $r_{\rm out}$ to change its sonic state from subsonic to supersonic and continues to proceed. Meanwhile, centrifugal repulsion becomes comparable against gravity that causes the accumulation of matter around the black hole. Due to this, a virtual barrier is developed that eventually triggers the discontinuous transition of the flow in the subsonic branch as a shock wave. After the shock, flow velocity gradually increases and again becomes supersonic at $r_{\rm in}$ before falling into the black hole. For shock, the following shock conditions \citep{Taub1948} are needed to be satisfied which are given by,
$$\begin{aligned}
&[\rho u^r]=0, \qquad [(e+p)u^tu^r]=0,\\
&{\rm and}\quad[(e+p)u^ru^r + pg^{rr}]=0,\\
\end{aligned}\eqno(21)$$
where quantities within the square brackets denote their differences across the shock front. In this work, we denote the location of the shock transition as $r_s$ which measures the size of the post-shock corona ($i.e.$, PSC). It may be noted that solutions of this kind that passes through $r_{\rm out}$, $r_s$ and $r_{\rm in}$ successively are known as global transonic shocked accretion solutions.

In Fig. \ref{fig:fig03}, we compare two typical accretion solutions corresponding to REoS and IEoS where both solutions harbor shock waves. In Fig. \ref{fig:fig03}a, we depict the variation of Mach number ($M$) with radial distance ($r$) where the input parameters are chosen as ${\cal E}=1.001$, $\lambda=1.98$ and $a_k=0.99$, respectively and subsonic flow is injected from $r=10^{3}$ with $\Gamma=1.4896$. Results plotted using solid and dotted curves are for REoS and IEoS, respectively. For REoS, the flow becomes supersonic after passing through the outer critical point at $r_{\rm out}=210.3884$ and continues its journey towards the black hole. Meanwhile, inflowing matter starts experiencing centrifugal repulsion and eventually encounters discontinuous shock transition in the subsonic branch at $r_s=14.5090$ as the shock conditions are satisfied there. This is indicated by the solid vertical arrow. Just after the shock transition, flow momentarily slows down and then gradually picks up its radial velocity due to gravitational attraction. Ultimately, the flow enters into the black hole supersonically after crossing the inner critical point at $r_{\rm in}=1.4696$. Similar to REoS, we observe the flow to pass through the shock (shown by the dotted arrow) for IEoS also, however, the outer critical point, shock, and inner critical point are obtained at different locations as $r_{\rm out}=215.5590$, $r_{s}=120.5611$ and $r_{\rm in}=1.4132$, respectively. In Fig. \ref{fig:fig03}b, we show the profile of adiabatic index ($\Gamma$) as function of $r$. As expected, $\Gamma$ remains constant (=1.4869) all throughout for IEoS whereas it decreases as the flow accretes towards the black hole for REoS. In general, accreting matter is compressed due to shock transition and consequently, density of the flow shoots up across the shock front which is measured by defining the compression ratio as $R=\sigma_+/\sigma_-$, where $\sigma=\rho H$ and `$+$' and `$-$' signs denote quantities calculated at immediate post-shock and pre-shock region. 
In the case of the above two solutions for REoS and IEoS, $R$ is calculated as $2.91$ and $1.22$, respectively. It may be noted that the location of the shock renders the size of the post-shock corona (PSC) where the soft photons from the pre-shock flow interact with the hot electrons of PSC via inverse Comptonization process to produce hard radiations \citep[and references therein]{Nandi-etal2018}. Thus, according to our model, both shock location and compression ratio seem to play a decisive role in 
determining the spectral features of the accretion disc around the black holes \citep{Chakrabarti-Titarchuk1995,Mandal-Chakrabarti2005b}. Since the accretion disc structures calculated using REoS and IEoS are not in agreement and REoS is developed based on the  physically motivated formalism, we, therefore, point out that it would be appropriate to utilize the accretion solutions yielded from 
REoS to study the observable properties of the black hole sources.

\subsection{Parameter Space for Shock}

\begin{figure}
	\includegraphics[scale=0.425]{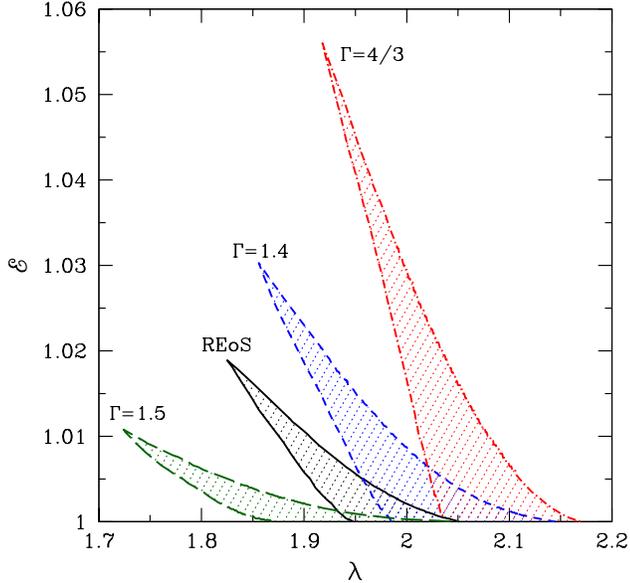}
	\caption{Parameter space for shock in $\lambda -{\cal E}$ plane. Region bounded with solid curve is for REoS and the same bounded by dot-dashed ($\Gamma=4/3$), dashed ($\Gamma=1.4$) and long-dashed ($\Gamma=1.5$) are for IEoS. Here, we choose $a_k =0.99$. See text for details.
	}
	\label{fig:fig04}
\end{figure}

In this section, we examine the range of the flow parameters, namely energy (${\cal E}$) and angular momentum ($\lambda$) that admits shock induced global accretion solutions around the rotating black holes. In Fig. \ref{fig:fig04}, we identify the effective domain of the parameter space for the shock in the $\lambda - {\cal E}$ plane where the region bounded by the solid curve is obtained for REoS. In addition, we also separate the shock parameter space for IEoS where dot-dashed, dashed and long-dashed boundaries denote the results for $\Gamma = 4/3, 1.4$ and $1.5$, respectively. In all the cases, we choose $a_k=0.99$. Clearly, noticeable disagreement is seen among the shock parameter spaces obtained for REoS and IEoSs. 

\begin{figure}
	\includegraphics[scale=0.75]{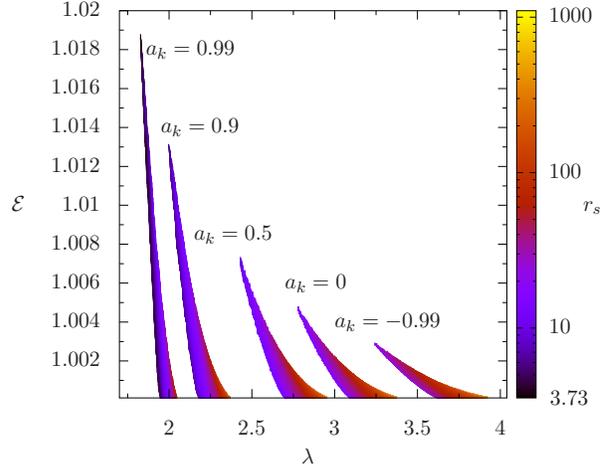}
	\caption{Two-dimensional projection of the three-dimensional plot of angular momentum ($\lambda$), energy (${\cal E}$) and shock location ($r_s$) for $a_k=0.99$, $0.9$, $0.5$, $0$ and $-0.99$ (left to right). In the right, vertical color coded bar denotes the range of $r_s$ calculated for REoS. See text for details.
	}
	\label{fig:fig05}
\end{figure}

We continue to study the shock parameter space around rotating black hole having different spin values. In Fig. \ref{fig:fig05}, we present the shock parameter space for $a_k = 0.99, 0.9, 0.5, 0.0$ and $-0.99$ (left to right), respectively where two-dimensional projection of the three-dimensional plot spanned with $\lambda$, ${\cal E}$ and $r_s$ is shown. In the figure, vertical color coded bar in the right side refers the range of $r_s$ calculated using REoS where we obtain the minimum value of shock location as $r^{\rm min}_{s} = 3.7314$ and the maximum value of shock location as $r^{\rm max}_{s} = 1071.5519$. We observe that the effective bounded region of the shock parameter space gradually shifts towards the lower angular momentum side as the spin of the black hole $a_k$ is increased. These findings are in agreement with the results of \cite{Aktar-etal2015} and \cite{Kumar_Chattopadhyay2017}. 
From the figure, it is clear that shocks generally form close to the horizon for the rapidly rotating black holes ($a_k=0.99$). Moreover, for a given $a_k$ and $\lambda$, shocks can also form at smaller radii when ${\cal E}$ is decreased.
On the other hand, for fixed $a_k$ and ${\cal E}$, shocks, in general, settle down at larger radii for flows with higher $\lambda$. This clearly  infers that centrifugal repulsion seems to play a crucial role in deciding the shock transition in the accretion disc.

\begin{figure}
\includegraphics[scale=0.75]{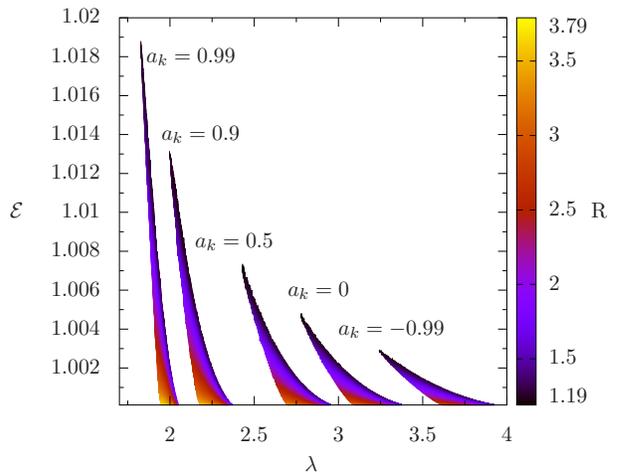}
\caption{Two-dimensional projection of the three-dimensional plot of angular momentum ($\lambda$), energy (${\cal E}$) and compression ratio ($R$) for $a_k=0.99, 0.9$, $0.5$, $0$ and $-0.99$ (left to right). In the right, vertical color coded bar denotes the range of $R$ calculated for REoS. See text for details.
}
\label{fig:fig06}
\end{figure}

Next, we demonstrate the shock parameter space in $\lambda - {\cal E}$ plane in terms of compression ratio ($R$) calculated using REoS. As before, in Fig. \ref{fig:fig06}, we examine the two-dimensional projection of the three-dimensional plot of $\lambda$, ${\cal E}$ and $R$, respectively for $a_k=0.99, 0.9, 0.5, 0$ and $-0.99$ (left to right) where vertical color code indicates the range of $R$, where we find the minimum value of $R$ as $R^{\rm min}=1.19$ and the maximum value of $R$ as $R^{\rm max}=3.79$, respectively. Note that the effective region of the shock parameter spaces displayed here are exactly identical to Fig. \ref{fig:fig05}. We observe that for a given $a_k$ and ${\cal E}$, inflowing matter experiences weak compression ($i.e$, $R \rightarrow 1$) when the angular momentum ($\lambda$) is relatively high and {\it vice versa}. On the other hand, for a given $a_k$ and $\lambda$, $R \rightarrow 1$ for flows with high ${\cal E}$ and {\it vice versa}. Overall, upon comparing Fig. \ref{fig:fig05} and Fig. \ref{fig:fig06}, it appears that $r_s$ and $R$ are closely related for flows accreting around the rotating black holes.

\subsection{$r_s - R$ correlation and QPO frequency ($\nu_{{\rm QPO}}$)}

\begin{figure*}
	\includegraphics[width=0.45\textwidth]{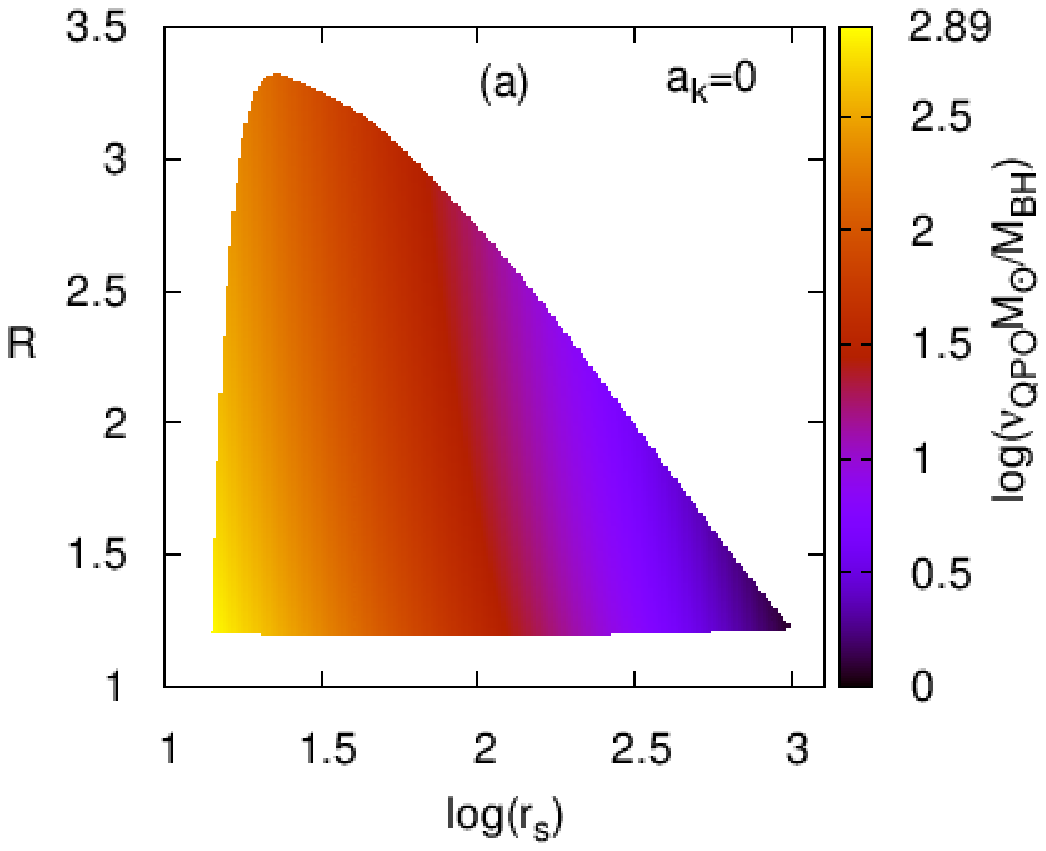}
	\includegraphics[width=0.45\textwidth]{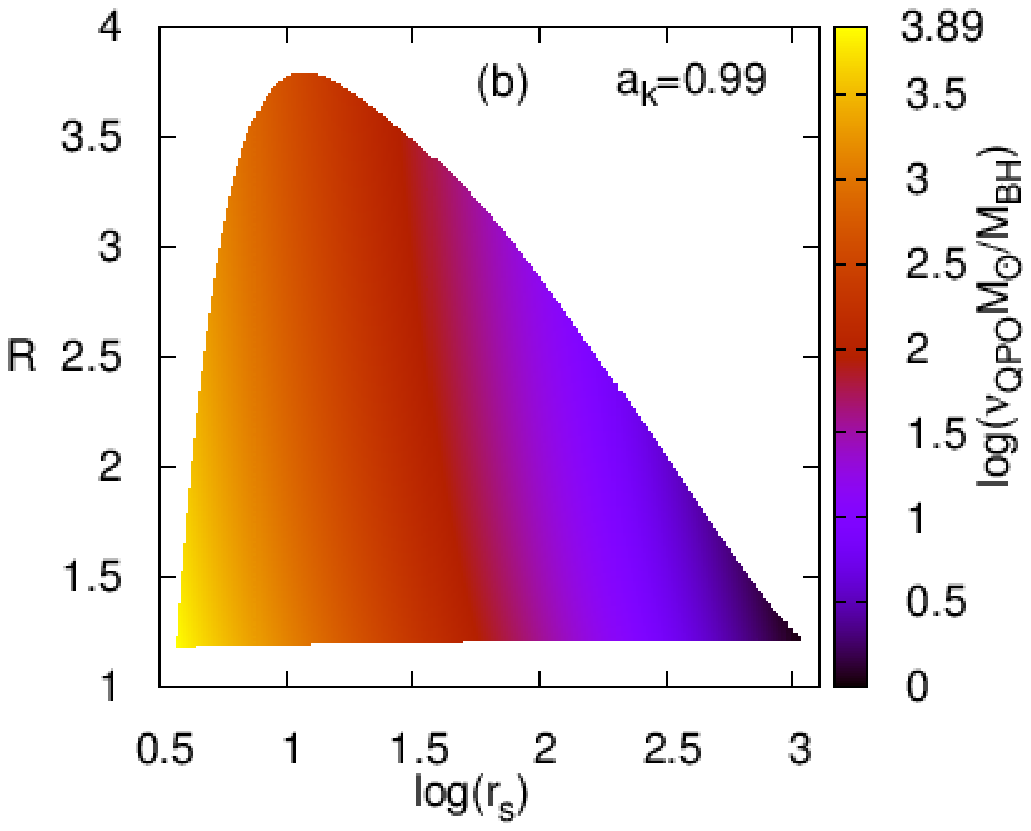}
	\caption{Two-dimensional projection of three-dimensional plot of \{$r_s, R, \nu_{\rm QPO} (M_\odot/M_{\rm BH}) \}$ where vertical color code bar indicates the range of QPO frequency in logarithmic scale. We choose $a_k = 0.0$ (a) and $a_k=0.99$ (b). See text for details.
	}
	\label{fig:fig07}
\end{figure*}

In this section, we intend to explore an important aspect of shock waves in an accretion disc. We have already shown that accreting matter passes through the standing shock wave provided the shock conditions are satisfied. 
Interestingly, when the shock conditions are not favorable, but the entropy of the flow at the inner critical point ($r_{\rm in}$) is higher than the outer critical point ($r_{\rm out}$), shock front exhibits non-steady behavior. This happens because of either resonance oscillation where infall time scale becomes comparable to the post-shock cooling time scale \citep{Molteni-etal96} or dynamical oscillations where the flow viscosity above its critical limit triggers the unstable perturbation in the flow \citep{Das-etal14}. The nature of the shock oscillation generally yields as quasi-periodic, and the frequency of this quasi-periodic oscillation (QPO) \citep{Chakrabarti-Manickam00,Iyer-etal2015} is computed as,
$$
\nu_{\rm QPO} =\frac{c/r_g}{\sqrt{2} \pi R r_s \sqrt{r_s - 2}},
\eqno(22)
$$
where $r_g (=2GM_{BH}/c^2)$ denotes the Schwarzschild radius. Now, for a given set of input parameters, namely \{$\lambda, {\cal E}, a_k$\}, $r_s$ and $R$ are uniquely determined and upon employing these values in equation (22), it is straightforward to estimate $\nu_{\rm QPO}$. Following this, we therefore retrace the shock parameter space in $r_s - R$ plane in lieu of $\lambda-{\cal E}$ plane and in Fig \ref{fig:fig07}, we display the two-dimensional projection of the three-dimensional plot of $\log (r_s)$, $R$ and $\log (\nu_{\rm QPO})$.
In Fig. \ref{fig:fig07}a, results are obtained for $a_k=0.0$ where we find $r_s^{\rm max}=999.3955$. Similarly, in Fig. \ref{fig:fig07}b, we choose $a_k=0.99$ and obtain $r_s^{\rm max}=1071.5519$. For $M_{\rm BH}=10M_\odot$, we get $\nu^{\rm max}_{\rm QPO}=77.68$ Hz and $783.50$ Hz corresponding to $a_k=0.0$ and $a_k=0.99$, respectively.
In both panels, vertical color coded bar indicates the range of $\nu_{QPO}$ in logarithmic scale. We observe that accretion flow exhibits high frequency QPOs when shock forms close to the horizon and {\it visa versa} irrespective to the black hole spin $a_k$. Interestingly, since the minimum value of the shock radius ($r^{\rm min}_{s}$) decreases with $a_k$, the maximum QPO frequency is ascertained around extremely rotating black hole ($a_k=0.99$).

It may be noted that in this work, we focus only on the axisymmetric shock oscillation model to explain the observed single peak HFQPO features. However, in reality, the shock can be non-axisymmetric as well that yields the spiral shock transition in the accretion flow. In fact, the shock transition between two-armed to three-armed spirals may be potentially viable to explain the 2:3 frequency ratio \citep{Chakrabarti-Wiita1993,Chakrabarti-2009} as observed in some of the BH sources \citep{Belloni-etal2012,Motta-2016}. However, the study of the non-axisymmetric behavior of the accretion flow around the black hole is beyond the scope of present work.

\section{Astrophysical Implications}

\begin{figure}
	\includegraphics[scale=0.4]{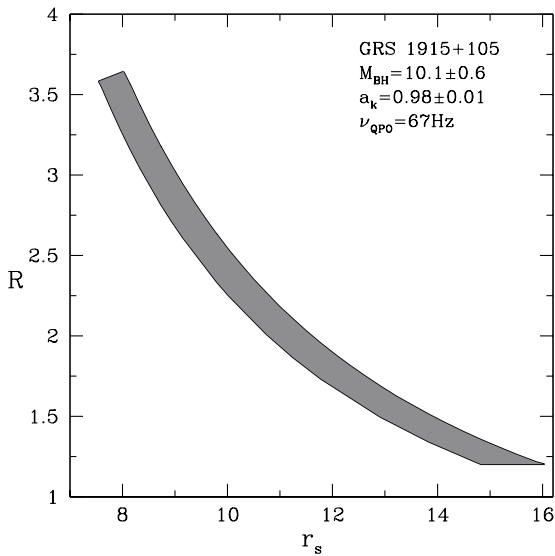}
	\caption{Variation of compression ratio ($R$) as function of shock location ($r_s$) that yields $\nu_{QPO} = 67$ Hz for Galactic black hole source $GRS~1915+105$. The ranges of $R$ and $r_s$ are obtained as the consequences of the mass and spin values as $9.5 \le M_{BH}/M_\odot \le 10.7$, and $0.97 \le a_k \le 0.99$, respectively. See text for details.}
	\label{fig:fig08}
\end{figure}

Recent observations confirm the significant detection of HFQPOs signature in few transient as well as persistent black hole sources  \citep{Altamirano-Belloni2012,Belloni-etal2012,Belloni-Altamirano2013}.
In this section, we apply our formalism to explain the plausible origin of the HFQPO and its connection with the spin of the black hole sources. In doing so, we choose two well studied Galactic black hole sources, namely $GRS~1915+105$ and $GRO~J1655-40$, respectively and carry out the analysis in this section.

First, we consider the persistent black hole source $GRS~1915+105$ as its mass and spin are well constrained. Recently, \citet{Steeghs-etal2013} estimated its mass as $M_{BH}=10.1\pm0.6 M_\odot $ and \citet{Miller-etal2013} reported its spin value as $a_k= 0.98\pm0.01$. Using these fundamental properties of the source, it is straight forward to calculate the frequency of QPO ($\nu_{QPO}$) from equation (22) for a given set of ($r_s,~R$). Since this source is known to exhibit high frequency QPOs ($\sim 67$ Hz \citet{Morgan-etal1997,Belloni-Altamirano2013}, we self-consistently compute $r_s - R$ correlation that yields $\nu_{QPO} = 67$ Hz. In Fig. 8, we show the variation of $R$ as function of $r_s$. In the figure, we consider the range of both mass and spin parameters as $9.5 \le M_{BH}/M_\odot \le 10.7$, and $0.97 \le a_k \le 0.99$ and the obtained results are displayed by the shaded region bounded by the solid curves. It is clear from the figure that HFQPOs observed in $GRS~1915+105$ can be understood using the present model provided shocks form close to the black hole ($7.5 \lesssim r_s \lesssim 16$) having compression ratio ranging from weak ($R \sim 1.2$) to strong ($R \sim 3.6$) limit.

Next, we choose the transient black hole source $GRO~J1655-40$. Several attempts have been made to obtain the precise measurements of mass and spin of $GRO J1655-40$. Using timing analysis, \citet{Motta-etal2014} measured the black hole mass and spin as $M_{BH}= 5.31 \pm 0.07 M_\odot$ and $a_k=0.290\pm0.003$, respectively. \citet{Stuchlik-kolos2016} estimated the black hole mass in the range $5.1 < M_{BH}/M_\odot < 5.5$ and spin $a_k < 0.3$. Meanwhile, \citet{Shafee-etal2006} calculated the spin of the black hole by spectral modeling in the range $0.65 < a_k <0.75$. On the other hand, \citet{Aktar-etal2017}, recently constrained the range of the spin parameter as $a_k \ge 0.57$ by modeling the oscillation of post-shock corona. Although mass of this source is well constrained, however, all these studies indicate contradictory claims particularly in the context of spin measurement of the black hole source. To address this issue, we employ our formalism to calculate the HFQPO at $\nu_{QPO} = 450$ Hz which has been observed in $GRO~J1655-40$ \citep{Remillard-etal1999,Strohmayer2001,Aktar-etal2017}. Generally, this particular HFQPO is seen when the source resides in the anomalous state with high flux values as well as hardness ratio of $0.3-0.8$ \citep{Belloni-etal2012}. While computing the frequency of HFQPO ($\nu_{QPO}$), we choose $M_{\rm BH} = 5.31~M_\odot$ and freely vary all the input parameters $(\lambda, {\cal E})$ including spin ($a_k$) of the black hole. With this, we find that $GRO~J1655-40$ can exhibit HFQPO at $450$ Hz provided its spin parameter $a_k\ge 0.85$ with $r_s \le 6.9$ and $R \ge1.22$.

\section{Conclusions}

Using the general relativistic approach, we study the shock induced global accretion solutions around Kerr black holes and compare the results obtained for the relativistic equation of state (REoS) as well as the ideal equation of state (IEoS). Accretion solutions of this kind are potentially viable as they can explain the observable properties of the black hole candidates. The main results of this work are summarized as follows.

(1) To the best of our knowledge, for the first time we find that there exists an upper limit of outer critical point location ($r^{\rm max}_{\rm out}$) for REoS whereas $r^{\rm max}_{\rm out}$ continues to remain unbound for IEoS. Since shock ($r_s$) forms in between $r_{\rm in}$ and $r_{\rm out}$ ($i.e.$, $r_{\rm in} < r_s < r_{\rm out}$), the maximum shock radius ($r^{\rm max}_{s}$) must be lower than $r^{\rm max}_{\rm out}$ and consequently, we observe an upper bound of $r^{\rm max}_{\rm out}$ as well for REoS (see Fig. \ref{fig:fig01}). 

(2) Considering both REoS and IEoS, we calculate the global transonic shocked accretion solutions for flows having identical input parameters around a black hole of spin $a_k = 0.99$ and find that solutions differ significantly (see Fig. \ref{fig:fig03}). Further, we identify the range of parameters in $\lambda-{\cal E}$ plane that admits shock for REoS and IEoS (see Fig. \ref{fig:fig04}). Here again, noticeable disagreement is seen. Since REoS satisfactorily describes the realistic accretion flow, we put emphasis on the accretion solutions yielded from REoS in order to study the observable properties of the black hole sources.

(3) In order to fit and compare the observed spectrum of the black hole sources, two component advective flow (TCAF) solutions are used as a local model in {\texttt {XSPEC}} software of HEAsoft \citep{Debnath-etal2014,Iyer-etal2015} where shock location ($r_s$) and compression ratio ($R$) are treated as free model parameters. To check consistency, in this work, we examine the two-dimensional projection of three-dimensional plot of $\{\lambda, {\cal E}, r_s \}$ (see Fig. \ref{fig:fig05}) and $\{\lambda, {\cal E}, R \}$ (see Fig. \ref{fig:fig06}) as function of $a_k$. We find that for a given set of the input parameters \{$\lambda, {\cal E}, a_k$\}, $r_s$ and $R$ are determined uniquely and therefore, {\it we argue that these two quantities ($r_s$ and $R$) must not be chosen arbitrarily while fitting the observed spectrum of the black hole sources}. It may be noted that the ranges of $r_s$ and $R$ are obtained from our model calculation as: for $a_k=0.0$: $14.1569 \le r_s \le 999.3955$ and  $1.19 \le R \le 3.31$ and for $a_k=0.99$: $3.7314 \le r_s \le 1071.5519$ and  $1.19 \le R \le 3.79$.

(4) We specify that when shock conditions are favorable, shock front is expected to start exhibiting non-steady behavior which is quasi-periodic in nature. We phenomenologically estimate the frequency of this quasi-periodic oscillation ($\nu_{\rm QPO}$) of the shock front and retrace the shock parameter space in $r_s -R$ plane instead of $\lambda-{\cal E}$ plane where two-dimensional projection of three-dimensional plot of $\{r_s, R, \nu_{QPO}\}$ is displayed (see Fig. \ref{fig:fig07}). We find that when shock oscillation takes place close to the horizon, it exhibits high frequency QPO. As shock can form very close to the horizon for the rapidly rotating black holes, in this work, we find the maximum QPO frequency as $\nu^{\rm max}_{\rm QPO}=783.50$ Hz for $a_k=0.99$ and $M_{\rm BH} = 10M_\odot$ (see Fig. \ref{fig:fig07}).

(5) Finally, we employ our formalism to understand the plausible origin of the HFQPO and its linkage with the spin parameter considering two well studied Galactic black hole sources, namely $GRS~1915+105$ and $GRO~J1655-40$, respectively. As the mass and spin of $GRS~1915+105$ are well constrained, we use these fundamental parameters to study the $r_s - R$ correlation that yields the HFQPO at $\nu_{QPO}= 67$ Hz. On the contrary, since the spin parameter of $GRO~J1655-40$ remains an unsettled issue, we use our formalism to constrain spin of this source. We find that HFQPO at $\nu_{QPO} = 450$ Hz can be explained in $GRO~J1655-40$ provided it spins very rapidly with $a_k \ge 0.85$. This result is consistent with some of the earlier findings \citep{Sramkova-etal2015,Aktar-etal2017}.

At the end, we point out that the present work is carried out considering some approximations. We ignore the effect of dissipations, namely viscosity, radiative processes, magnetic fields, etc. We also do not take into account of mass loss from the accretion disc. Although the implementation of all these issues is beyond the scope of the present work, however, the basic conclusion of this work is expected to remain unaltered due to the above approximations.

\section*{Acknowledgments}

 Authors thank the anonymous reviewer for providing useful comments that significantly strengthened the manuscript. AN thanks GD, SAG; DD, PDMSA and Director, URSC for encouragement and continuous support to carry out this research.

\appendix
\newpage 

\section{Typical Example of angular momentum variation}
  
  \begin{figure}
    \centering
  	\includegraphics[scale=0.45]{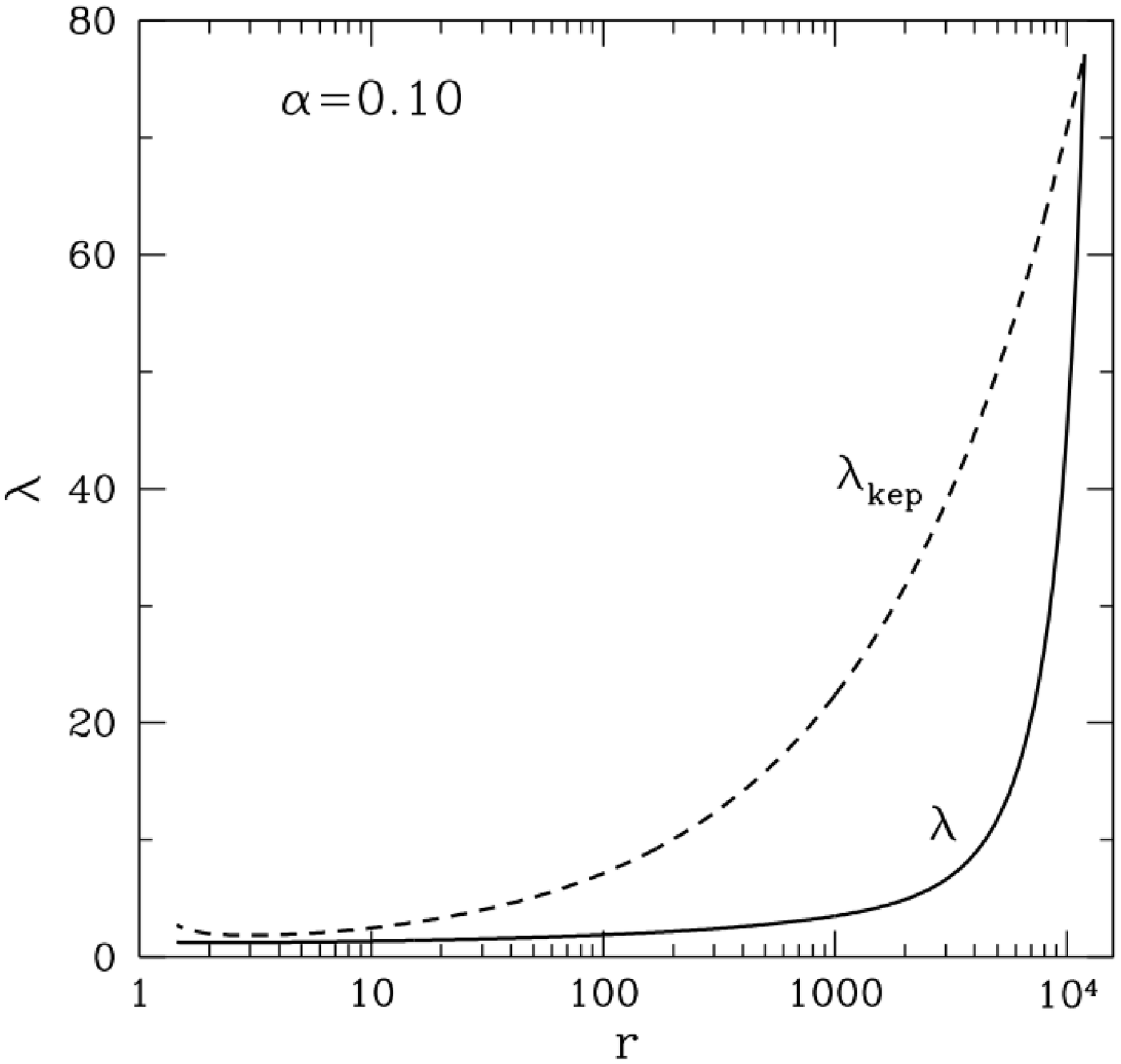}
  	\caption{Variation of angular momentum ($\lambda$) of the accretion flow with radial coordinate ($r$) as shown by solid curve. Results depicted using dashed curve represent the Keplerian angular momentum distribution (${\lambda_{\rm Kep}}$). Here, we use a unit system as $2G = M_{\rm BH} = c =1.$ See text for details given in appendix A.
  	}
  \end{figure}
  
 In an accretion disc around a black hole, the presence of viscosity is ubiquitous. However, the viscous time scale generally exceeds the infall time scale of the accreting matter at the inner part of the disc and therefore, matter does not get enough time to transport angular momentum outward due to the differential motion leaving the flow to be inviscid in nature \citep{Fukue1987,Chakrabarti1989}. This particularly happens for accretion flow possessing multiple critical points. In Fig. A1, a typical example is depicted where we compare the angular momentum distribution ($\lambda$) for a viscous accretion solution \citep{Chakrabarti-Das2004} with the Keplerian angular momentum ($\lambda_{\rm Kep}$). In the figure, we inject accreting matter from the outer edge of the disc ($r_{\rm edge}$) with flow variables as $r_{\rm edge} =11890.8$, angular momentum $\lambda_{\rm edge} = \lambda_{\rm Kep}$, velocity $v_{\rm edge} = 0.00064$, sound speed $a_{\rm edge} = 0.00688$ and viscosity parameter $\alpha =0.1$, respectively. Fig. A1 clearly indicates that the angular momentum variation remains quite insensitive for wide range of inner radial coordinate. The above findings support the assumption of inviscid nature of accretion flow around the black hole.

\section{Calculation of $\frac{\lowercase{dv}}{\lowercase{dr}}\big|_{\rm \lowercase{c}}$}
The gradient of radial velocity at the critical point given by,
$$
\frac{dv}{dr}\bigg|_{c}=\frac{N_1-D_2\pm\sqrt{(N_1-D_2)^2+4D_1N_2}}{2D_1},
\eqno(A1).
$$
where,
$$\begin{aligned}
N_1 = &\frac{2a_s^2}{\Gamma + 1}(N_{11} + N_{12})A'\Theta_{11},\\
N_2 = &N_{21} + N_{22} + N_{23} + N_{24} + N_{25} + N_{26},\\
D_1 = &\gamma_v^2\bigg[1 + \frac{2a_s^2}{\Gamma + 1}\bigg\lbrace\frac{1}{v^2} 
- \frac{A'\Theta_{22}}{v}\bigg\rbrace\bigg],\\
D_2 = &-\frac{2a_s^2\gamma_v^2A'}{(\Gamma + 1)v}\Theta_{11},\\
N_{11}= & \frac{\left(r-a_k^2\right)}{r\Delta} + \frac{5}{2r},
N_{12}= -\frac{1}{2\mathcal{F}}\frac{d\mathcal{F}}{dr},
N_{21}= \frac{2 (r-1)}{(r-2)^2 r^2},\\
N_{22}= &-\frac{4 a_K\lambda \gamma _{\phi }^2}{r^3 \Delta}
-\frac{2a_k\lambda \gamma _{\phi }^2\Delta'}{r^2 \Delta^2}
+\frac{4 a_k\lambda \gamma _{\phi } \gamma _{\phi }'}{r^2 \Delta},\\
N_{23}= & -\frac{8 a_k^2 \gamma _{\phi}^2}{(r-2) r^3 \Delta}
-\frac{4 a_k^2 \gamma _{\phi }^2\Delta' }{(r-2) r^2 \Delta^2}
+\frac{8 a_k^2 \gamma _{\phi }\gamma _{\phi }'}{(r-2) r^2 \Delta}
-\frac{4 a_k^2 \gamma _{\phi }^2}{(r-2)^2 r^2 \Delta },\\
N_{24}=&\Omega \gamma _{\phi }^2\lambda   \frac{2 a_k^2-(r-3) r^2\Delta'}{r^2 \Delta^2}
-\gamma _{\phi }^2\lambda \frac{2 a_k^2-(r-3) r^2 \Omega'}{r^2 \Delta}\\
&-2 \lambda\Omega  \gamma _{\phi }\frac{2 a_k^2-(r-3) r^2 \gamma _{\phi }'}{r^2 \Delta}
+2 \lambda \Omega \gamma _{\phi }^2 \frac{2 a_k^2-(r-3) r^2 }{r^3 \Delta}\\
&+\lambda\Omega  \gamma _{\phi }^2\frac{  r^2+2 (r-3) r }{r^2 \Delta},\\
N_{25}=&\frac{2 a_k \left(2 a_k^2-(r-3) r^2\right) \Omega  \gamma_{\phi}^2 \Delta '}
{(r-2) r^2 \Delta ^2}-\frac{2 a_k \left(2 a_k^2-(r-3) r^2\right) \gamma_{\phi }^2 
\Omega '}{(r-2) r^2 \Delta }\\
&-\frac{4 a_k \left(2 a_k^2-(r-3) r^2\right) \Omega  \gamma_{\phi } 
\gamma _{\phi }'}{(r-2) r^2 \Delta }+\frac{2 a_k \left(2 a_k^2-(r-3) r^2\right) 
\Omega  \gamma_{\phi }^2}{(r-2)^2 r^2 \Delta }\\
&+\frac{4 a_k \left(2 a_k^2-(r-3) r^2\right) \Omega  \gamma _{\phi }^2}
{(r-2) r^3 \Delta }-\frac{2 a_k \left(-r^2-2 (r-3) r\right) \Omega  
\gamma _{\phi }^2}{(r-2) r^2 \Delta },\\
N_{26}=&\frac{2a_s^2}{\Gamma + 1}\bigg[N_{111} +
 N_{121} + (N_{11} + N_{12})A'\Theta_{11}\bigg],\\
N_{111}=&  -\frac{r-a_k^2}{r^2 \Delta}-\frac{\left(r-a_k^2\right)\Delta'}
{r \Delta^2}-\frac{5}{2 r^2}+\frac{1}{r \Delta},\\
N_{121}=& -4 a_k^2 r \frac{\left(a_k^2+r^2\right) 
\Delta '-4 r \Delta}{\left(a_k^2+r^2\right)^4-4 a_k^4 \Delta^2},\\
A'=&\frac{1}{\Theta} + \frac{\Gamma'}{\Gamma} -\frac{\Gamma'}{\Gamma + 1} 
- \frac{a_s^2(\Gamma + 1)}{\Gamma\Theta},\\
\Theta_{11}=&-\frac{2\Theta}{(N + 1)}\bigg[\frac{\left(r-a^2\right)}{r\Delta} 
+ \frac{5}{2r}- \frac{1}{2\mathcal{F}}\frac{d\mathcal{F}}{dr}\bigg],\\
\Theta_{22}=&-\frac{2\Theta\gamma_v^2}{(N + 1)v},
\Omega = \frac{2 a_k+\lambda  (r-2)}{a_k^2 (r+2)-2 a_k \lambda +r^3}\\
\Gamma'=&\frac{\partial \Gamma}{\partial \Theta}{~~\rm  and~}
\gamma_\phi'=\frac{\gamma_\phi^3}{2}\lambda\Omega'.\\
\end{aligned}$$
Here, all the quantities have their usual meaning.

\bsp 
\label{lastpage}
\end{document}